\documentclass{PoS}

\usepackage{bm}

\title{Leptonic widths of heavy quarkonia: QCD/NRQCD matching for the
electromagnetic current at $O(\alpha_s v^2)$}

\ShortTitle{Leptonic widths of heavy quarkonia}

\author{A. Hart\\
        School of Physics, University of Edinburgh,
            King's Buildings, Edinburgh EH9 3JZ, U.K.
        \email{hart@ph.ed.ac.uk}}
\author{\speaker{G.M. von Hippel}\\
        Department of Physics, University of Regina,
            Regina, SK, S4S 0A2, Canada
        \email{vonhippg@uregina.ca}}
\author{R.R. Horgan\\
        DAMTP, CMS, University of Cambridge,
            Cambridge CB3 0WA, U.K.
        \email{R.R.Horgan@damtp.cam.ac.uk}}

\abstract{
          We construct the S-wave part of the electromagnetic vector
          annihilation current to $\mathcal{O}(\alpha_s v^2)$, where
          $v$ is the non-relativistic quark velocity, for heavy quarks
          whose dynamics are described by the NRQCD action on the
          lattice. The NRQCD vector current for $Q\bar{Q}$
          annihilation is expressed as a linear combination of lattice
          operators with quantum numbers $L=0$, $J^P=1^-$, and the
          coefficients are determined by  matching to the
          corresponding continuum  current in QCD to
          $\mathcal{O}(v^2)$ at one-loop. The annihilation 
          channel gives a complex amplitude with Coulomb-exchange and
          infrared singularities, making a careful choice for
          the contours of integration and infrared subtraction
          functions in the numerical integration necessary. An
          automated vertex generation program written in Python is
          employed, allowing us to use a realistic NRQCD action and an
          improved gluon lattice action; a change in the definition of
          either action is easily accommodated in this procedure. The
          final result is applicable to simulations of electromagnetic
          decays of heavy quarkonia, notably the $\Upsilon$ meson.
}

\FullConference{XXIVth International Symposium on Lattice Field Theory\\
                July 23-28, 2006\\
                Tucson, Arizona, USA}

\begin{document}


\section{Introduction}
\label{intro}

Leptonic widths of heavy quarkonia such as the $\Upsilon$ or the
$J/\psi$ are an important test of electroweak Standard Model in the
heavy quark sector: The heavy quarks are the heaviest Standard Model
particles and hence should be sensitive to possible new physics at or
above the electroweak scale, and leptonic decays have experimentally
clean signatures. Moreover, ratios of leptonic widths can be measured
to good accuracy both experimentally and on the lattice.

The CLEO collaboration has experimental results to few-percent
precision \cite{Rosner:2005eu}:
\begin{equation}
       \frac{\Gamma_{\Upsilon(2S)\to e^+e^-}}%
            {\Gamma_{\Upsilon(1S)\to e^+e^-}} = 0.457(6)
\end{equation} 
which has to be compared with the current best lattice result
\cite{Gray:2005ur}
\begin{equation}
       \frac{\Gamma_{\Upsilon(2S)\to e^+e^-}M^2_{\Upsilon(2S)}}%
            {\Gamma_{\Upsilon(1S)\to e^+e^-}M^2_{\Upsilon(1S)}} = 0.48(5)
\end{equation}  

There is thus a challenge to the lattice community to obtain a
precision on theoretical predictions that can be compared to that
achieved experimentally.


\section{ Matching S-wave decays between NRQCD and QCD}
\label{mswbnaq1}

The leptonic width of a $\bar{Q}Q$ state is given by
\begin{equation}
       \Gamma_{\bar{Q}Q\to l^+l^-} = \frac{8\pi}{3 M_{\bar{Q}Q}}
       \left|\left<0\left|\mathbf{J}^{QCD}\right|\bar{Q}Q\right>\right|^2
       e_Q^2 \alpha_{em}^2
\end{equation} 
with all the nonperturbative QCD contributions encapsulated in the
matrix element $\left<0\left|\mathbf{J}^{QCD}\right|\bar{Q}Q\right>$.
Unfortunately, it is not possible to simulate QCD with heavy quarks
directly due to their short Compton wavelengths, so Non-Relativistic
QCD (NRQCD) has to be used in lattice simulations of heavy quarkonia.

Hence, we need to match the desired QCD matrix element to a series of
NRQCD matrix elements which can be measured on the lattice:
\begin{equation}
       \left<0\left|\mathbf{J}^{QCD}\right|\bar{Q}Q\right> = 
       \sum_i a_i
       \left<0\left|\mathbf{J}_i^{NRQCD}\right|\bar{Q}Q\right>
\end{equation}
where the $a_i$ are the matching coefficients which we need to
determine. For the case of S-wave decays, which we will study
exclusively in this paper, we can take the NRQCD currents to be
$\mathbf{J}_i^{NRQCD} = {\bm \sigma} \left(\frac{\Delta^2}{M^2}\right)^i$.

To compute the matching coefficients perturbatively, we expand both the
coefficients and the matrix elements perturbatively:
\begin{equation}
       a_i = \sum_n \alpha_s^n a_i^{(n)} \hskip2cm
       \left<0\left|\mathbf{J}\right|\bar{Q}Q\right> =
       \sum_n \alpha_s^n \left<0\left|\mathbf{J}\right|\bar{Q}Q\right>^{(n)}
\end{equation} 
and match order by order in $\alpha_s$.

In the $\Upsilon$ system, the order of the NRQCD expansion parameters
is $v^2\sim\alpha_s\sim 10\%$. \textit{Prima facie}, this would
suggest that to achieve $\sim 1\%$ accuracy, we would need to go to
$\mathcal{O}(\alpha_s^2,\alpha_s v^2,v^4)$. However, in matrix element
ratios the $\mathcal{O}(\alpha_s^2)$ terms cancel, and hence we only
need to include $\mathcal{O}(\alpha_s,\alpha_s v^2,v^4)$ corrections
for $\sim 1\%$ accuracy.

If we are only interested in the ratio of leptonic widths of, say,
$\Upsilon(2s)$ and $\Upsilon(1s)$, we do not care about the overall
normalisation of the matrix element, and so for each decay we need
only consider instead the quantity
\begin{equation}
\frac{M_\textrm{ME}}{a_0} = \left< J_0 \right> + 
\frac{a_1}{a_0} \left< J_1 \right> + 
\frac{a_2}{a_0} \left< J_2 \right> \; .
\end{equation}
and we can define matching coefficients for the ratio as
\begin{eqnarray}
b_1 & \equiv &
\frac{a_1}{a_0} = \frac{a_1^{(0)}}{a_0^{(0)}} +
\frac{\alpha_s}{a_0^{(0)}} \left[
a_1^{(1)} - \frac{a_1^{(0)} a_0^{(1)}}{a_0^{(0)}}
\right]  \; ,
\nonumber \\
b_2 & \equiv &
\frac{a_2}{a_0} = \frac{a_2^{(0)}}{a_0^{(0)}} \; .
\end{eqnarray}
In the following, we work in the Breit frame, where the decaying meson is
stationary and the quark has momentum $p^\mu=(iE,0,0,M v)$, use $v$ as
the non-relativistic expansion parameter (which is exact at the order
to which we are working) and treat the quarks as being exactly
on-shell (which can also be shown to be justified).


\section{ The improved NRQCD action }
\label{tina1}

The improved NRQCD action used for simulations is
\begin{equation}
            \mathcal{S}_{NRQCD} = \sum_{x,t} \psi^\dagger \psi -
	    \psi^\dagger
	    \left( 1 - \frac{a\delta H}{2} \right) 
	    \left( 1 - \frac{aH_0}{2n} \right)^n 
	    U_4^\dagger 
	    \left( 1 - \frac{aH_0}{2n} \right)^n
	    \left( 1 - \frac{a \delta H}{2} \right)
	    \psi
\end{equation} 
with
\newcommand{\delsq}{\Delta^{(2)}}
\newcommand{\delfour}{{\Delta^{(4)}}}
\newcommand{\Mbz}{{(aM)}}
\newcommand{\delv}{\bm{\nabla}}
\newcommand{\delvt}{\tilde{\bm{\nabla}}}
\newcommand{\Ev}{\tilde{\bm{E}}}
\newcommand{\Bv}{\tilde{\bm{B}}}
\newcommand{\sigmav}{\bm{\sigma}}
\begin{eqnarray*}
              a H_0 & = & \frac{\delsq}{2M} \\
              a \delta H & = & 
	      - c_1 \, \frac{(\delsq)^2}{8\Mbz^3}
	      + c_2\,\frac{i}{8\Mbz^2}\left(\delv\cdot\Ev -
                                            \Ev\cdot\delv\right)
	      - c_3\,\frac{1}{8\Mbz^2} \sigmav\cdot(\delvt\times\Ev -
                                                    \Ev\times\delvt) \\
              && 
	      - c_4\,\frac{1}{2\Mbz}\,\sigmav\cdot\Bv
	      + c_5\,\frac{\delfour}{24\Mbz}
	      - c_6\,\frac{(\delsq)^2}{16n\Mbz^2}
\end{eqnarray*} 
where $n$ is a stability parameter for the euclidean-space
Schr\"odinger equation, which must satisfy $n\ge 3/(Ma)$ for numerical
stability. To the perturbative order considered here, we can take
$c_i=1$.

As our glue action, we use a Symanzik improved action with tadpole
improved links.


\section{ Automatically generating Feynman rules }
\label{agfr1}

In order to correctly determine the desired matching coefficients, we
need to consider exactly the same NRQCD action in perturbation theory
as is used in simulations. For the improved NRQCD action, the Feynman
rules become extremely complicated: The QQg vertex, e.g., has $\sim
8,000$ terms, and the QQgg vertex has $\sim 70,000$ terms! It is clear
that a traditional manual treatment would be extremely cumbersome and
error-prone.

For this reason, we have developed HiPPy, an automated tool for
generating Feynman rules from lattice actions. HiPPy is written
entirely in Python with companion modules in Fortran 95, and is freely
available from any of the authors. The main strength of HiPPy lies in
its great flexibility: HiPPy is capable of handling not only various
kinds of NRQCD actions, but also relativistic (staggered, Wilson
\dots) quark and gluon actions with or without improvement. A
description of HiPPy has been published in \cite{Hart:2004bd}, and it
is currently being used by HPQCD member to calculate a variety of
different improvement and renormalisation constants. Due to its
flexible design, a HiPPy-based program can easily accommodate a change
in the quark or gluon action being used without the need for changes
to the user code.


\section{ Matching at tree level }
\label{matl1}

At tree-level, the relevant matrix elements are given by
\begin{eqnarray*}
       \left<0\left|\mathbf{J}^{QCD}\right|\bar{Q}Q\right>^{(0)} &=&
       \bar{v}(-\mathbf{p}) {\bm \gamma} u(\mathbf{p}) =
       \chi^\dag{\bm \sigma}
       \left(\frac{2}{3}+\frac{M}{3E}\right)\psi \\
       \left<0\left|\mathbf{J}_i^{NRQCD}\right|\bar{Q}Q\right>^{(0)} &=&
       g_i(v) \chi^\dag{\bm \sigma}\psi
\end{eqnarray*} 
where
\begin{eqnarray*}
       g_0(v) &=& 1 \\
       g_1(v) &=& -\frac{4}{(Ma)^2}\sin^2\left(\frac{aMv}{2}\right) \\
       g_2(v) &=& \frac{4}{(Ma)^2}\left[4\sin^2\left(\frac{aMv}{2}\right)-\sin^2(aMv)\right]
\end{eqnarray*} 
Expanding these matrix elements in powers of $v^2$, we determine
$a_i^{(0)}$ to match:
\begin{equation}
       a_0^{(0)} = 1 \;\;\;\;\;\;\;\; a_1^{(0)} = \frac{1}{6} 
       \;\;\;\;\;\;\;\; a_2^{(0)} = \frac{1}{8}-\frac{(aM)^2}{72}
\end{equation}


\section{ Matching to one-loop order }
\label{mtolo1}

Expanding the matching condition to first order in $\alpha_s$ gives
\begin{equation}
     \sum_i\underbrace{a_i^{(1)}}\limits_{\textrm{wanted}}
     \overbrace{\left<0\left|\mathbf{J}_i^{NRQCD}\right|\bar{Q}Q\right>^{(0)}}
     \limits^{\textrm{{known} functions of $v$}}  =
     \underbrace{\left<0\left|\mathbf{J}^{QCD}\right|\bar{Q}Q\right>^{(1)}}
     \limits_{I_{QCD}} -
     \underbrace{\sum_i a_i^{(0)}
     \left<0\left|\mathbf{J}_i^{NRQCD}\right|\bar{Q}Q\right>^{(1)}}
     \limits_{I_{NRQCD}}
\end{equation} 
Both the QCD and the NRQCD matrix elements on the right-hand side
contain odd powers of $v$ coming from the Coulomb-exchange
singularity; however, only even powers of $v$ are available for
matching on the left-hand side, so the odd powers must cancel exactly.

In fact, the odd powers of $v$ are a purely infrared phenomenon, and
are known exactly:
\begin{equation}
       I_{odd} = \frac{h(v)}{12v} = -\Im \left\{\frac{4}{3} \int
       \frac{d^4k}{(2\pi)^4} \frac{h(v)}{(\bm{k}^2 + \mu^2)
       (ik_0 - \frac{\bm{k}^2 + 2\bm{k} \cdot \bm{p}}{2M})
       ( ik_0 + \frac{\bm{k}^2 + 2\bm{k} \cdot \bm{p}}{2M})} \right\}
\label{eqn:oddpowers}
\end{equation}
where $h(v)$ is a known even function of $v$. We can hence
analytically subtract the odd powers from both QCD and NRQCD by
rearranging the right-hand side as
\begin{equation}
       I_{QCD} - I_{NRQCD} = (I_{QCD} - I_{odd}) -
       \left.(I_{NRQCD} - I_{odd})\right|_{\mathcal{B}} +
       \left.I_{odd}\right|_{\mathbf{R}^4\setminus\mathcal{B}}
\end{equation} 
where $\mathcal{B}$ signifies integration over the Brillouin zone
only. 

The term $(I_{QCD} - I_{odd})$ is known analytically, while the other
terms are calculated numerically using farmed VEGAS on the CCHPCF
SunFire Galaxy class computer.

The results obtained at various $v$ are then fitted with the ansatz
\begin{equation}
       I_{QCD} - I_{NRQCD} = a_0^{(1)} - a_1^{(1)} g_1(v)
\end{equation} 
to obtain the matching coefficients at one-loop order.


\subsection{ The QCD form factors }
\label{tqff1}

The relevant QCD on-shell form factors 
\begin{eqnarray*}
       \left<0\left|\mathbf{J}^{QCD}\right|\bar{Q}Q\right>^{(1)} &=&
         F_{1,R}^{(1)}(4E^2)\bar{v}(-\mathbf{p}){\bm\gamma}u(\mathbf{p}) +
         F_2^{(1)}(4E^2)\bar{v}(-\mathbf{p})\frac{\tilde\mathbf{q}}{2M}u(\mathbf{p})\\ 
         &=&
       \Big(F_{1,R}^{(1)}(4E^2)f_1(v^2) + F_2^{(1)}(4E^2)f_2(v^2) \Big)
       \chi^\dag{\bm\sigma}\psi
\end{eqnarray*} 
are simply the corresponding QED form factors (times a colour factor),
which are well known at the one loop level. The $F_2$ term is both UV-
and IR-finite; the $F_1$ term is UV-finite by virtue of the Ward
identity, but IR divergences (which cancel against those in the NRQCD
matrix elements) remain. Moreover, the $F_1$ term contains odd powers
of $v$ which arise from the $1/v$ Coulomb-exchange singularity. As
mentioned before, these odd powers are known to be the same in QCD and
NRQCD, allowing them to be subtracted analytically.


\subsection{ The NRQCD self-energy }
\label{tnse1}

To account for the wavefunction renormalisation in NRQCD, as well as
to establish the connection between the renormalised mass in terms of
which the QCD form factors are formulated and the bare mass appearing
in the NRQCD action, we need to compute the self-energy of the NRQCD
heavy quarks.

The NRQCD self-energy, which is the sum of the usual ``rainbow'' and
``tadpole'' diagrams, can be decomposed as
\begin{equation}
       a\Sigma(p_0,\bm{p}) =  A + B(p_0,\bm{p}) \; aT(\bm{p})
       +  C(p_0,\bm{p}) \left[ 1 - e^{-iap_0} \left(1 - aT(\bm{p})
       \right) \right]
\end{equation} 
where $T(\bm{p})$ is the tree-level kinetic energy, and from this form
it is straightforward to derive the needed quantities, namely the
wavefunction and (kinetic) mass renormalisation constants as
\begin{eqnarray*}
         Z_\psi(\bm{p}) & = & 1 + \alpha_s \left( a\Sigma + \frac{\partial
         a\Sigma}{\partial (iap_0)} \right)_\textrm{on-shell} \\
         Z_M & = & 1 + \alpha_s 2M \left. \frac{d a\Sigma}{d \bm{p}^2}
         \right|_{\bm{p^2}=0}
\end{eqnarray*} 
Given the complicated nature of the Feynman rules, we employ
automatic differentiation techniques \cite{vonHippel:2005dh} to
calculate the derivatives.


\subsection{ The NRQCD vertex correction }
\label{tnvc1}

\begin{figure}
\includegraphics[width=\textwidth,keepaspectratio=,clip=]{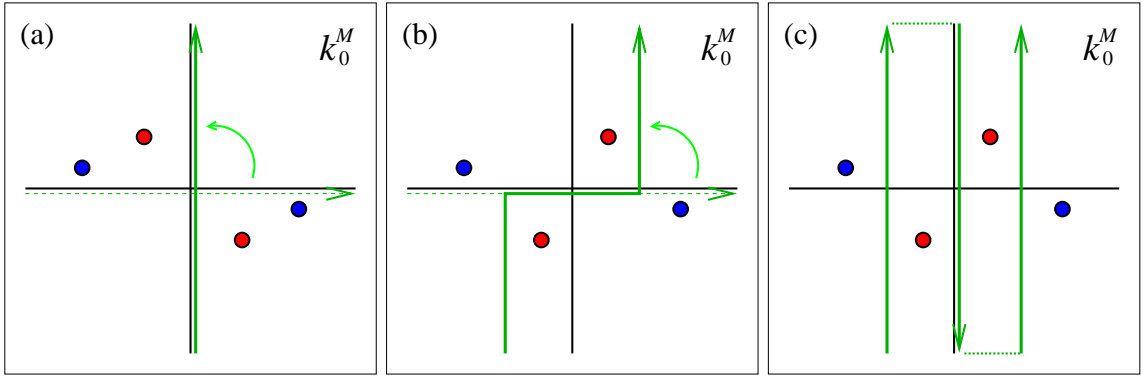}
\caption{The position of the quark (red) and gluon (blue) poles in the
  Minkowskified energy plane. A normal Wick rotation (a) is only
  possible for $\mathbf{k}^2>-2\mathbf{p\cdot k}$; otherwise, the
  integration contour has to be deformed as in (b); for numerical
  work, the equivalent contour shown in (c) is adopted instead.}
\label{fig:k0plane}
\end{figure}

The NRQCD vertex correction suffers from the same infrared divergences
that appear in the corresponding QCD diagram; we use a small gluon
mass $\mu$ as an infrared regulator.

In terms of the Minkowskified energy variable $k_0^M$, the poles of
the integrand (in the continuum limit) are at
$\pm k_0^M =  \sqrt{\mathbf{k}^2+\mu^2} - i\epsilon$ for the gluons,
and $\pm k_0^M = \left(\frac{2\mathbf{p\cdot k}+\mathbf{k}^2}{2M}\right)
- i\epsilon$ for the quarks, as shown in fig. \ref{fig:k0plane}.
Hence, a normal Wick rotation between Euclidean and Minkowskian
momenta is possible only for $\mathbf{k}^2>-2\mathbf{p\cdot k}$, since
otherwise, the quark poles cross imaginary axis. We therefore need to
deform the Euclidean contour of integration to avoid the quark poles
and pick up the correct analytic continuation to Minkowksi space, and
the choice of contour is shown in  fig. \ref{fig:k0plane} (c).

To subtract the odd powers of $v$ coming from the Coulomb-exchange
singularity, we use the integral form of eqn. \ref{eqn:oddpowers}.
The evaluation of the resulting finite integral is still quite hard
numerically, and takes up the major part of the computer time used.
 
For $i>0$, the matrix elements of the NRQCD current $\mathbf{J}_i$
also contains tadpole-type diagrams. Since each tadpole loop reduces
the $v$-dependence of the result by one power of $v^2$, this leads to
a contamination of the lower-order matching coefficients by ''mixing
down'', which would appear to necessitate a complete recalculation if
higher-order currents are added in later. The solution lies in
defining subtracted currents $\bar{J}_i = z_{ij} J_j^{NRQCD}$
where  $z_{ij}$ is defined such that we have
$\left<0\left|\bar{J}_i\right|\bar{Q}Q\right>^{(n)} =
\mathcal{O}(v^{2i}) $ for all $n$. For details, we have to refer the
reader to \cite{Hart:2006ij}.


\section{ Results }
\label{results}

\begin{figure}
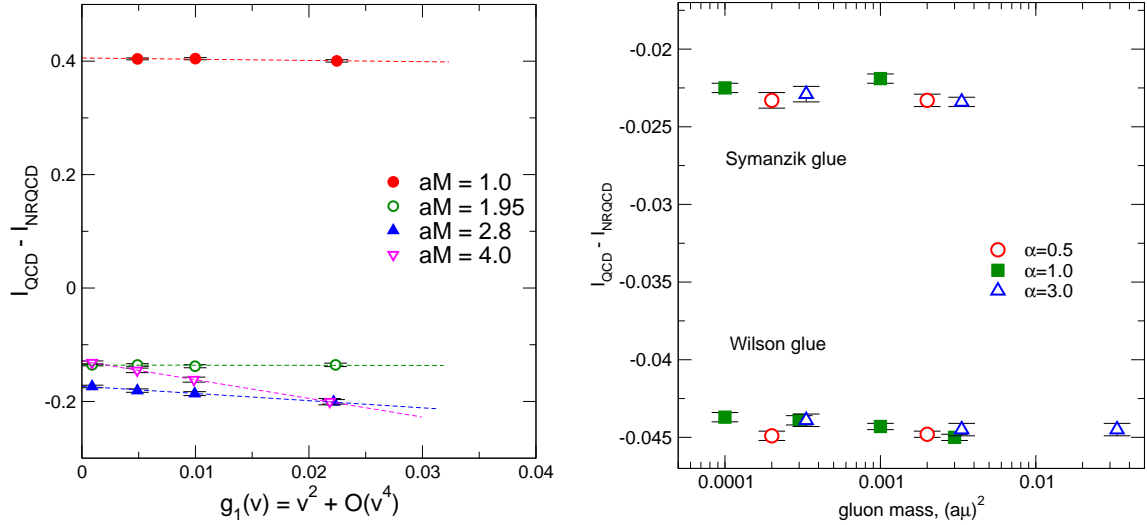

\includegraphics[width=0.48\textwidth,keepaspectratio=,clip=]{full_extrap.eps}
\hfill
\includegraphics[width=0.49\textwidth,keepaspectratio=,clip=]{jnw_total.eps}
\caption{Left: the numerical results with the fit to extract the
  matching coefficients; right: a plot of results in different gauges
  against the infrared gluon mass, showing gauge and gluon mass
  independence.}
\label{fig:results}
\end{figure}

\begin{table}[t]
\begin{center}
\begin{tabular}{llllll}\hline\hline
$M_0a$ & $n$ & $a_0^{1}$   & $a_1^{1}$ & $b_1^{1}$ & $b_2^{0}$ \\\hline
4.0    &  2  & -0.1288(27) & -3.32(29) & -3.30(30) & -0.0972 \\
2.8    &  2  & -0.1732(21) & -1.35(22) & -1.32(22) &  0.0161 \\
1.95   &  2  & -0.1358(16) &  0.26(17) &  0.14(17) &  0.0722 \\
1.0    &  4  &  0.4056(20) & -0.50(17) & -0.56(17) &  0.1111 \\\hline\hline
\end{tabular}
\end{center}
\caption{The matching coefficients, as a function of the 
\textit{bare} heavy quark mass, for the leptonic width ($a_i$) 
and leptonic width ratio ($b_i$). Note that 
$a_0^{(0)} = 1$, $a_1^{(0)} = b_1^{(0)} = \frac{1}{6}$, and that 
there is \textit{no} subtraction to prevent mixing down.}
\label{tab:results}
\end{table}

We have run our calculation at a number of different quark masses
corresponding to the bottom quark on the MILC supercoarse, coarse and
fine ensembles, and to the charm quark on the super-coarse
ensemble. We have also performed extensive tests of gauge invariance,
infrared regulator independence, and agreement with known results for
$a_0^{(1)}$ at $v=0$ in the case of simpler NRQCD actions. A plot of
our results can be seen in fig. \ref{fig:results}, as can be a plot
showing the gauge and regulator independence of our results. Our final
results for the matching coefficients are given in
tab. \ref{tab:results}.

The authors thank G.P. Lepage and C.T.H. Davies for useful discussions
and the Cambridge--Cranfield High Performance Computing Facility for
advice and use of the SunFire supercomputer.
This research was supported in part by the U.K. Royal Society, 
the Canadian Natural Sciences and Engineering Research Council (NSERC)
and the Government of Saskatchewan.


\begin{thebibliography}{99}

\bibitem{Rosner:2005eu}
{\bf CLEO} Collaboration, J.~L. Rosner {\em et~al.},
  {\em Phys. Rev. Lett.} {\bf 96}
  (2006) 092003, [\href{http://arxiv.org/abs/hep-ex/0512056}{{\tt
  hep-ex/0512056}}].

\bibitem{Gray:2005ur}
A.~Gray {\em et~al.},
  {\em Phys. Rev.} {\bf D72} (2005) 094507,
  [\href{http://arxiv.org/abs/hep-lat/0507013}{{\tt hep-lat/0507013}}].

\bibitem{Hart:2004bd}
A.~Hart, G.~M. von Hippel, R.~R. Horgan, and L.~C. Storoni,
  {\em J.
  Comput. Phys.} {\bf 209} (2005) 340--353,
  [\href{http://arxiv.org/abs/hep-lat/0411026}{{\tt hep-lat/0411026}}].

\bibitem{vonHippel:2005dh}
G.~M. von Hippel,
  {\em Comput. Phys. Commun.} {\bf 174} (2006)
  569--576, [\href{http://arxiv.org/abs/physics/0506222}{{\tt
  physics/0506222}}].

\bibitem{Hart:2006ij}
A.~Hart, G.~M. von Hippel, and R.~R. Horgan,
  \href{http://arxiv.org/abs/hep-lat/0605007}{{\tt hep-lat/0605007}}.

\end{thebibliography}
\end{document}